\documentclass[conference]{IEEEtran}
\IEEEoverridecommandlockouts
\usepackage{cite}
\usepackage{amsmath,amssymb,amsfonts}
\usepackage{algorithmic}
\usepackage{graphicx}
\usepackage{textcomp}
\usepackage{xcolor}
\def\BibTeX{{\rm B\kern-.05em{\sc i\kern-.025em b}\kern-.08em
    T\kern-.1667em\lower.7ex\hbox{E}\kern-.125emX}}
    
\usepackage{booktabs} 
\usepackage{multirow}
\usepackage{booktabs}
\usepackage{graphicx}  
\usepackage{diagbox}
\usepackage[braket]{qcircuit} 
\usepackage{threeparttable} 
\usepackage{stfloats} 
\usepackage{enumitem} 
\usepackage{fancyheadings} 

\usepackage{listings}
\usepackage{color}

\usepackage{times}
\usepackage{graphicx}
\usepackage{epsf}
\usepackage{verbatim}
\usepackage{url}
\usepackage{color}
\usepackage{alltt}

\newcommand{\Caption}{\caption}

\newcommand{\CodeIn}[1]{{\small\texttt{#1}}}

\newcommand{\Comment}[1]{}

\newenvironment{CodeOut}{\begin{scriptsize}}{\end{scriptsize}}

\newcommand{\CenterCell}[1]{\multicolumn{1}{c|}{#1}}

\newcommand{\SmallSpace}{\vspace*{-1.4ex}}

\begin{document}

\title{Identifying Bug Patterns in Quantum Programs\\
}

\author{\IEEEauthorblockN{Pengzhan Zhao}
\IEEEauthorblockA{\textit{Kyushu University}\\
zhao.pengzhan.813@s.kyushu-u.ac.jp}
\and
\IEEEauthorblockN{Jianjun Zhao}
\IEEEauthorblockA{\textit{Kyushu University}\\
zhao@ait.kyushu-u.ac.jp}
\and
\IEEEauthorblockN{Lei Ma}
\IEEEauthorblockA{\textit{Kyushu University}\\
malei@ait.kyushu-u.ac.jp}
}

\maketitle
\begin{abstract}
Bug patterns are erroneous code idioms or bad coding practices that have been proved to fail time and time again, which are usually caused by the misunderstanding of a programming language's features, the use of erroneous design patterns, or simple mistakes sharing common behaviors. 
This paper identifies and categorizes some bug patterns in the quantum programming language Qiskit and briefly discusses how to eliminate or prevent those bug patterns. We take this research as the first step to provide an underlying basis for debugging and testing quantum programs.
\end{abstract}


\section{Introduction}
Debugging and testing are critical parts of an integrated software development method in modern software development. An appropriate method of bug finding can quickly help developers locate and fix bugs. A software bug is regarded as the abnormal program behaviors which deviate from its specification~\cite{allen2002bug}, including poor performance when a threshold level of performance is included as part of the specification. Bug patterns are recurring relationships between potential bugs and explicit errors in a program; they are common coding practices that share similar symptoms and have been proven to fail time and time again. Bug patterns are usually raised from the misunderstanding of language features, the misuse of positive design patterns, or simple mistakes having common behaviors. Bug patterns are an essential complement to the traditional design patterns~\cite{gamma1995design}, just as a good programmer needs to know design patterns that can improve the software design quality in various aspects, also to be a good software developer, the knowledge of common causes of faults is a need in order to know how to fix the software bugs.

Quantum programming is the process of designing and building executable quantum computer programs to achieve a particular computing result and is drawing increasing attention recently. A number of quantum programming approaches are available to write quantum programs, for instance, Qiskit~\cite{ibm2017qiskit}, Q\#~\cite{svore2018q}, ProjectQ~\cite{projectq2017projectq}, Scaffold~\cite{abhari2012scaffold}, and Quipper\cite{green2013quipper}. However, the current research so far in quantum programming is focused on problem analysis, language design, and implementation. Even though program debugging and software testing are important, they have received little attention in the quantum programming paradigm~\cite{zhao2020quantum}. The new complexity introduced in quantum programming makes it difficult to find the bugs in the quantum source code. Until now, only a few approaches have been proposed for testing and debugging quantum software~\cite{li2020projection,liu2020quantum,huang2019statistical,honarvar2020property,miranskyy2019testing,miranskyy2020your,ali2021assessing,wang2018quanfuzz} and none of the previous work has focused on the bug pattern identification in a quantum programming language. The testing and debugging issues remain a big problem for quantum programs~\cite{zhao2020quantum}.

We may not know what types of bugs are unique or common happened to quantum programs without a proper bug pattern classification, and this poses several restrictions on the research and development of programs in the language.

\begin{itemize}[leftmargin=2em]
\setlength{\itemsep}{3pt}
  \item Developers may not know what type of bugs are most likely to happen in a program, and therefore do not know how to prevent them. In other words, a programmer would lack a piece of fundamental knowledge on how to write bug-free code.
  \item Testers do not have sufficient knowledge of how to write adequacy test cases that can effectively cover the most common potential errors. Only when having an idea of how the common bugs happened in programs can the tester set up criteria for better addressing the specific bugs.
  \item Software maintenance staff do not know which language features are more likely to result in the incorrect code, so they cannot clearly view the current system when doing the maintenance tasks.
\end{itemize}

The bug patterns presented in this paper may help to solve these problems mentioned above. Identifying such patterns in a quantum programming language can help programmers improve their productivity in finding bugs and reduce software maintenance costs. Bug pattern identification can also help language designers and tool developers develop efficient bug-finding techniques to locate bugs in quantum programs' source code through program analysis.

Furthermore, the information on bug patterns provides a basis for further research on debugging quantum programs. It provides insight into the possible consequences of different bug types and summarizes the common behaviors among similar ones. Also, it can be used to recognize faults that have been already existed and prevent potential bugs. The bug patterns identified for the quantum programming language Qiskit could be seen as the first step towards studying general bug patterns for quantum programming languages.

This paper chooses the widely used quantum programming language Qiskit as our target language and identifies the common bug patterns in Qiskit programs. We also provide an example for each bug pattern to illustrate the pattern's symptoms. To the best of our knowledge, our work presented in this paper is the first attempt to identify some bug patterns existing in Qiskit programs systematically.

The rest of the paper is organized as follows. Section~\ref{sec:background} briefly introduces the background knowledge of quantum programming in Qiskit and the error-prone features introduced by quantum programs. Section~\ref{sec:bug-pattern} describes the identified bug patterns in Qiskit in detail. Related work is discussed in section~\ref{sec:related-work}, and concluding remarks are given in Section~\ref{sec:conclusion}.

\section{Background}
\label{sec:background}
We next briefly introduce the background information on programming in Qiskit and the error-prone features in Qiskit programs.

\subsection{Qiskit}

Qiskit is one of the most widely used open-source frameworks for quantum computing, allowing us to create algorithms for quantum computers~\cite{koch2019introduction}. As a Python package, it provides tools for creating and manipulating quantum programs and running on prototype quantum devices and simulators~\cite{aleksandrowicz2019qiskit} and can use built-in modules for noise characterization and circuit optimization to reduce the impact of noise. Qiskit also provides a library of quantum algorithms for machine learning, optimization, and chemistry.

In Qiskit, an experiment is defined by a quantum object data structure that contains configuration information and the experiment sequences. The object can be used to get status information and retrieve results~\cite{mckay2018qiskit}.
Figure~\ref{fig:mesh1} shows a simple Qiskit program that illustrates the entire workflow of a quantum program. 
The function \CodeIn{Aer.get\_backend('qasm\_simulator')} returns a backend object for the given backend name(\CodeIn{qasm\_simulator}). 
The \CodeIn{backend} class is an interface to the simulator, and the actual name  of \CodeIn{Aer} for this class is \CodeIn{AerProvider}.
\noindent
After the experimental design is completed, the instructions are run through \CodeIn{execute} method.
The \CodeIn{shots} of the simulation, which means the number of times the circuit is run, is set to 1000 while the default is 1024.
When outputting the results of a measurement, the method \CodeIn{job.result()} is used to retrieve the measurement results. We can access the counts via the method \CodeIn{get\_counts(circuit)}, which gives the experiment's aggregate outcomes.


\subsection{Properties of Qubits}
In the following, we use Qiskit as an example to explain the characteristics of quantum bit (qubit) and the necessary execution process of a complete quantum program.

\begin{figure}[t]
  \begin{CodeOut}
\footnotesize{
  \begin{alltt}
    simulator = \textbf{Aer.get_backend}('qasm_simulator')

    qreg = \textbf{QuantumRegister}(3)
    creg = \textbf{ClassicalRegister}(3)
    circuit = \textbf{QuantumCircuit}(qreg, creg)

    circuit.\textbf{h}(0)
    circuit.\textbf{h}(2)
    circuit.\textbf{cx}(0, 1)
    circuit.\textbf{measure}([0,1,2], [0,1,2])
    job = \textbf{execute}(circuit, simulator, shots=1000)
    result = job.\textbf{result}()
    counts = result.\textbf{get_counts}(circuit)
    \textbf{print}(counts)
  \end{alltt}
}
\end{CodeOut}
\Caption{\label{fig:mesh1}A simple quantum program in Qiskit}
\end{figure}

In quantum computing, the basic unit of information is the quantum bit (qubit). As shown in Figure~\ref{fig:mesh1}, \CodeIn{qreg = QuantumRegister(3)} means assigning a quantum register of three qubits, and the value of each qubit is $\ket{0}$ by default. So the initial value of these three qubits is $\ket{000}$. Next, let the first and third qubits pass through the H (Hadamard) gate, as shown by \CodeIn{circuit.h(0)} and \CodeIn{circuit.h(2)}. In this way, the unique property {\it superposition} of qubits is realized, which means that each qubit can take on values of $\ket{0}$ and $\ket{1}$.
There is also an {\it entanglement} of qubit properties that only multiple qubits can achieve. The code in the sample program is \CodeIn{circuit.cx(0,1)}. That is to say, the first qubit is entangled with the second qubit through a CNOT (Controlled-NOT) gate operation. We measure the first qubit, and its output is \CodeIn{0} for 50 percent probability and \CodeIn{1} for 50 percent probability. After that, measuring the second qubit is 100 percent the same as the first measurement result.
Since the third qubit is not related to the first two qubits, the last qubit's measurement result is still taken with \CodeIn{0} for 50 percent probability and \CodeIn{1} for 50 percent probability.
The measurement statement of qubits shown in Figure~\ref{fig:mesh1} is \CodeIn{circuit.measure([0,1,2], [0,1,2])}. Measurement can lead to the collapse of a quantum superposition state to a classical state. There are many kinds of quantum measurements, and the projection measurement of a single qubit is used here. That is, each qubit is projected onto a state space consisting of base vectors $\ket{0}$ or $\ket{1}$. In this program, the final output is a three-bit array.

\subsection{Error-Prone Features in Qiskit Programs}
By focusing on the language features of Qiskit, we can classify the bug patterns in Qiskit into the following four categories.

\begin{itemize}[leftmargin=2em]\setlength{\itemsep}{3pt}
  \item {\bf Initialization}: A quantum program is a series of operations on qubits. The initial stage is to initialize the quantum registers to store the qubits that need to be manipulated. Then the classical registers are initialized to store the values of the measured qubits. This stage does not include setting the quantum state, as the quantum state setting needs to be implemented by a gate operation.
  Quantum registers and classical registers do not have to be of the equal initial size. When we use multiple classical bits to store the same qubit measurements, we need to initialize as many classical bits as possible. However, another case is that the initialized qubit is larger than the classical bit. Since the programmer does not intend to measure some qubits, it is assumed that there is no need to initialize the classical bits equal to the qubits. Nevertheless, this is also the reason why most programs go wrong.
  So a hasty initialization can cause some problems for subsequent programs.
  
  \item {\bf Gate Operation}: The core of quantum computing is to operate on qubits. Qiskit provides almost all the gates to implement algorithms in quantum programs~\cite{IBM}. To achieve the superposition of qubits, it must pass through the H (Hadamard) Gate. To achieve "entanglement" in the case of multiple qubits, it must pass through the CNOT ( controlled-NOT) gate. In quantum language, complex gate operations are decomposed into basic gates and gradually realized. Controlled gates are parameterized by two qubits, and double-controlled gates require three qubits. However, this does not mean that the double-controlled gate operates on three qubits at the same time. 
  Many errors may occur when inappropriately using gates that operate on the qubits multiple times. 
  
  \item {\bf Measurement}: When we want to obtain the output, we must perform a measurement operation on the target qubit. The measured qubit is returned as the classical state's value, which no longer has superposition properties. 
  So the qubit that has been measured cannot be used as a control qubit to entangle with other qubits.
  Although measurement is a simple operation, the program executing a measurement statement is very complicated. 
  It requires thousands of projection measurements of the qubits. Finally, it outputs all its possible results. Moreover, the number of occurrences of the result is used to obtain the size of the probability of outputting the correct value.
  Many errors start with the measurement statement because programmers do not really understand the effect of measurements on the state of qubits.
  
  \item {\bf Deallocation}: It is crucial to reset and release the qubits safely; otherwise, the auxiliary qubits in the entangled state will affect the output. Deallocation is not considered to be a specific operation due to the power of Qiskit. We do not need to reset the qubits manually.
  However, In some backends, not releasing all qubits can be problematic. 
  In Qiskit, not handling all the qubits in the entangled state can cause problems in the program or output unsatisfactory values.
\end{itemize}

For better understanding, we propose these bug patterns in terms of the quantum program execution order, which consists of four stages (processes) that the program's qubits go through, and each stage interacts with the others. 

\section{Bug Patterns in Qiskit}
\label{sec:bug-pattern}

We next introduce six bug patterns in Qiskit as examples. When introducing each bug pattern, we also show an example that contains this specific pattern. Since most bug patterns have some representation variants and alternatives, we choose the one that appears to be the most generally applicable. These bug patterns are also summarized in Table~\ref{table:catalog}. 

\subsection{Unequal Classical Bits and Qubits}

In Qiskit, each classical bit in the classical register stores a measured qubit value. Therefore, it is better to initialize quantum registers of the same size as classical registers.
Otherwise, the bug pattern of ``\textit{Unequal Classical Bits and Qubits}" may occur,  especially when the number of qubits in the quantum register is greater than that of classical bits in the classical register. From the point of view of program integrity, every used qubit should be measured.

As shown in Figure~\ref{fig:mesh2}, when we want to measure the third qubit, we receive an error message \textit{CircuitError: `Index out of range.'}
If we do not measure one of the qubits, then a qubit will not get reset.

\begin{figure}[t]
\begin{CodeOut}
\footnotesize{
\begin{alltt}
    qreg = \textbf{QuantumRegister}(3)
    creg = \textbf{ClassicalRegister}(2)
    circuit = \textbf{QuantumCircuit}(qreg, creg)
    
    circuit.\textbf{h}(0)
    circuit.\textbf{cx}(0, 1)
    circuit.\textbf{cx}(1, 2)
    circuit.\textbf{measure}([0,1,2], [0,1,2])
  \end{alltt}
}
\end{CodeOut}
\Caption{\label{fig:mesh2} Unequal classical bits and qubits}
\end{figure}

Another case is that the number of bits in the classic register is larger than the qubit. Unless we encounter the need to use multiple classical bits to store a qubit measurement, otherwise, this is not a good habit. On the one hand, resources are wasted when the program is actually developed, and on the other hand, outputting all classic bits will cause very messy results. Therefore we do not recommend this operation.

\subsection{Custom Gates not Recognised}
When defining a custom gate in a program, some programmers will want to define a basic gate that controls more than two qubits directly; the bug pattern \textit{Custom gates not recognised by Qiskit} may occur.
This pattern refers to a custom gate that does not use the gate class provided by Qiskit correctly. 
Alternatively, the gate is not recognized by Qiskit.

An example of an error code is shown in Figure~\ref{fig:mesh3}, which is a program that tends to define a three-qubit controlled gate. First define a gate named \CodeIn{my\_custom\_gate} using the \CodeIn{Gate} method, and control the number of qubits to three. When we call this gate, the program will have an error.
Because in \CodeIn{basic\_gates}, the custom gate \CodeIn{gt} is not the same as other Qiskit-based gates.

\begin{figure}[t]
\begin{CodeOut}
\footnotesize{
  \begin{alltt}
    qc = \textbf{QuantumCircuit}(3,3)
    gt = \textbf{Gate}('my_custom_gate', 3, [])
    
    qc.\textbf{h}(0)
    qc.\textbf{sdg}(0)
    qc.\textbf{y}(1)
    
    qc.\textbf{append}(gt, [0,1,2])
    qc.\textbf{add_calibration}(gt, [0,1,2], schedule)
    qc = \textbf{transpile}(qc, backend,
                   basis_gates=['u1','u2','cx', gt])
    qc.\textbf{measure}([0,1,2], [0,1,2])
  \end{alltt}
}
\end{CodeOut}
\Caption{\label{fig:mesh3} Custom gates not recognised by Qiskit}
\end{figure}

This bug pattern is mainly caused by programmers who do not really understand quantum gates. 
Quantum gates can only control a maximum of 2 qubits and are known as basic gates. The compound gates we usually use, such as the double-controlled gate CCX (Toffoli)~\cite{ibm2017qiskit}, are not the gates that directly control three qubits. 
Instead, multiple single-qubit gates and controlled gates are combined, resulting in a dual controlled gate effect.
The correct custom gate should be a composite gate combining the basic gates provided by Qiskit and applied to the circuit.

\subsection{Insufficient Initial Qubits}
When the \CodeIn{TwoLocal} method is used, a dual-local parametric circuit consisting of alternating rotating and  entangled layers can be formed. The two-local circuit is a parameterized circuit consisting of alternating rotation layers and entanglement layers. If the number of qubits of the variational form does not match. The bug Pattern ``\textit{Insufficient Initial Qubits}" may occur.

As shown in Figure~\ref{fig:mesh4}, which is part of the code for the Variational Quantum Eigensolver (VQE) algorithm.
When defining the VQE solver, method \CodeIn{TwoLocal} is used. As the \CodeIn{num\_qubits} is set to \CodeIn{1}.
In addition, the value of \CodeIn{num\_qubits} is replaced by any other value that does not match, the desired result is not obtained.
So it is important to initialize the values supported by the parameter \CodeIn{num\_quibits} when using parametric   circuits or methods involving quantum entanglement.  

\begin{figure}[t]
  \begin{CodeOut}
\footnotesize{
  \begin{alltt}
num\_qubits = 1
tl\_circuit = \textbf{TwoLocal}(num\_qubits, ['h', 'rx'], 'cz',
                      entanglement='full', reps=3, 
                      parameter\_prefix = 'y')
tl\_circuit.\textbf{draw}(output = 'mpl')
  \end{alltt}
}
\end{CodeOut}
\Caption{\label{fig:mesh4} Insufficient initial qubits}
\end{figure}

\subsection{Over Repeated Measurement} Some simulator backends are unable to execute the circuit when the measurement operation performed on the qubit is repeated too many times.
Alternatively, when some methods, such as \CodeIn{c\_if}, are called but do not give the correct result. This situation may lead to the bug pattern of ``\textit{Over Repeated Measurement}."

\begin{figure}[h]
  \begin{CodeOut}
\footnotesize{
  \begin{alltt}
\textbf{def} get_circuit(n):
    
    qreg = \textbf{QuantumRegister}(1)
    creg = \textbf{ClassicalRegister}(n)
    mreg = \textbf{QuantumRegister}(1)
    dreg = \textbf{ClassicalRegister}(1)
    circ = \textbf{QuantumCircuit}(qreg, mreg, creg, dreg)
        
    \textbf{for} i \textbf{in} range(n):
        circ.\textbf{measure}(qreg[0], creg[i])
        
    circ.\textbf{x}(mreg[0]).\textbf{c_if}(creg, 0)
    circ.\textbf{measure}(mreg[0], dreg[0])
    \textbf{return} circ
        
b_aer = \textbf{BasicAer.get_backend}('qasm_simulator')
aer = \textbf{Aer.get_backend}('qasm_simulator')
circ65 = \textbf{get_circuit}(65)
\textbf{print}("65clbits(Aer):",\textbf{execute}(circ65, aer).
      \textbf{result}().\textbf{get_counts}())
\textbf{print}("65clbits(Basic_Aer):",\textbf{execute}(circ65, b_aer).
      \textbf{result}().\textbf{get_counts}())
\end{alltt}
}
\end{CodeOut}
\Caption{\label{fig:mesh5} Over repeated measurement}
\end{figure}

To show this bug pattern, consider the piece of code in Figure~\ref{fig:mesh5}. This is a test used to measure quantum characteristics in a computing backend simulator repeatedly.
We use the Qiskit ``Aer" simulator backend and the Python-based quantum simulator module ``BasicAer" to simulate the circuit \CodeIn{qasm\_simulator}. The same qubit is used multiple times here. When we call \CodeIn{BasicAer}, the system may report an error that the number of qubits is greater than the maximum (24) for \CodeIn{qasm\_simulator}.
Not only that, the \CodeIn{c\_if} method we called did not get the desired result on the “Aer” backend simulator, that is, the qubits of the \CodeIn{mreg} register did not achieve flipping. While the code \CodeIn{circ.x(mreg[0]).c\_if(creg,0)} did not achieve.
And if \CodeIn{n=63} in the classic register \CodeIn{creg}, the system will hang.

In summary, we do not recommend excessive measurement operations on qubits. The measured qubit is placed in the first position of the quantum register, and then the measurement is placed in the second position. Such repeated operations are equivalent to operating ``N" multiple qubits. As a result, it can make the system extremely unstable.

\begin{figure}[b]
\begin{CodeOut}
\footnotesize{
  \begin{alltt}
    tq = \textbf{QuantumRegister}(3)
    tc0 = \textbf{ClassicalRegister}(1)
    tc1 = \textbf{ClassicalRegister}(1)
    tc2 = \textbf{ClassicalRegister}(1)
    teleport = \textbf{QuantumCircuit}(tq, tc0,tc1,tc2)
    
    teleport.\textbf{h}(tq[1])
    teleport.\textbf{cx}(tq[1], tq[2])
    teleport.\textbf{ry}(np.pi/4,tq[0])
    teleport.\textbf{cx}(tq[0], tq[1])
    teleport.\textbf{h}(tq[0])
    teleport.\textbf{barrier}()
    teleport.\textbf{measure}(tq[0], tc0[0])
    teleport.\textbf{measure}(tq[1], tc1[0])
    teleport.\textbf{cx}(tq[1], tq[2])
    teleport.\textbf{cz}(tq[0], tq[2])
    teleport.\textbf{measure}(tq[2], tc2[0])

    backend = Aer.\textbf{get_backend}('qasm_simulator')
    job = \textbf{execute}(teleport, backend, 
                  shots=1, memory=True).\textbf{result}()
    result = job.\textbf{get_memory}()[0]
    \textbf{print}(job.\textbf{get_memory}()[0]) 
  \end{alltt}
}
\end{CodeOut}
\Caption{\label{fig:mesh6} Incorrect operations after measurement}
\end{figure}

\subsection{Incorrect Operations after Measurement}

When the measurement is completed, we cannot use the measured qubit for entanglement. Otherwise, we will not get the desired result.
The result of the measurement can be treated as a classical value that no longer has the properties that the qubit has. If the measured value continues to be entangled with other qubits, which is used to change the target qubit state, it will be the bug pattern of ``\textit{Incorrect Operations after Measurement}." 

Considering the code snippet in Figure~\ref{fig:mesh6} taken from GitHub document~\cite{Qiskitters}, 
which realizes a quantum teleportation protocol. 
In the code, the last qubit's state should be changed according to the first two bits' measurement results. The wrong instructions in the example are \CodeIn{teleport.cx(tq[1],tq[2])} and \CodeIn{teleport.cz
(tq[0],tq[2])}, which entangle the measured qubit with the unmeasured qubit, and therefore affect the result of the last qubit. This mistake is quite common, and many programmers inadvertently use measured qubits. 
In this program, the correct code should be \CodeIn{teleport.z(tq[2]).c\_if(tc0,1)} as well as \CodeIn{teleport.x
(tq[2]).c\_if(tc1,1)}.

Although these erroneous operations follow quantum measurements, the reason for this lies in a poor understanding of the effect of measurement operations on qubit states.

\begin{table*}[t]
\caption{\label{table:catalog} A catalog of bug patterns in Qiskit}
\footnotesize{
\begin{center}
\begin {tabular} {|p{2.8cm}|p{0.8cm}|p{4cm}|p{4cm}|p{4cm}|}
\hline \textbf{Bug Patterns} &\CenterCell{\textbf{Category}}&\CenterCell{\textbf{Symptoms}}&\CenterCell{\textbf{Causes}}&\CenterCell{\textbf{Cures and Preventions}}\\

\hline Unequal Classical Bits and Qubits & 1 & Classical registers are not large enough to store the measured qubits & The initialized classical bits are smaller than the qubits used or to be measured & Try to initialize quantum and classical registers of the same size\\

\hline Custom gates not recognised & 2 & The program is unable to customize the gate function and will often report errors & Creating gates that directly control more than three qubits does not follow the principle of two qubit entanglement & Try to use the gates provided by Qiskit for the implementation of the algorithm\\

\hline Over Repeated Measurement & 3 & Output error or program error when measuring the same quantum bit multiple times with a \CodeIn{for} loop & number of measurements repeated several times & Reduction of meaningless measurements. \\

\hline Incorrect Operations after Measurement & 3 & Unable to get the desired post-measurement result & Continued manipulation of the qubit being measured, such as changing its state or re-entangling with other qubits & The measured result cannot be used as a condition unless it is re-operated and measured as the initial qubit after reset \\

\hline Unsafely Uncomputation & 4 & The program reports an error or does not achieve the desired result & Auxiliary qubits are not reset and remain entangled with the target qubit, which can affect the results of the target qubit measurement & Correctly reset or release all qubits to ensure they are in their initial or post-measurement states \\

\hline Insufficient Initial Qubits & 1 & Causes VQE not to respect the form of input variables and outputs the wrong circuit & When the \CodeIn{TwoLocal} method is used, a dual-local parametric circuit consisting of alternating rotating and entangled layers can be formed & When using parametric circuits or methods involving quantum "entanglement," initialize the values supported by the parameter \CodeIn{num\_qubits}. \\

\hline Inappropriately Modification of Register Size & 1 & Changing the register size may cause the program to report an error. Especially for building complex circuits & Changing the size of a register may change the hash value of the register and its bits, thus prohibiting it from being used as a key for structures such as sets & It is possible to reinitialize the registers. Otherwise, it is not recommended to modify the values of the registers without changing the variable names \\
\hline Method \CodeIn{measure\_all} &3& The program outputs the results of all measured qubits normally. However, it also outputs the classical register values & When the \CodeIn{measure\_all} method is used, the program automatically creates a classical register to store all the qubits being measured & If we want to call the \CodeIn{measure\_all} method to measure all qubits, we do not need to initialize the classical registers\\
\hline

\end{tabular}
\end{center}
}
\vspace*{-3ex}
\end{table*}

\subsection{Unsafely Uncomputation}
Qiskit is a compelling framework because it supports the automatic management of qubits, i.e., there is no need to do the work of unallocated qubits manually. 
However, as different program languages (e.g. Q\#) have their own implementations, which can lead to exceptions in different backends and the need to manually unallocated qubits. 
Considering other programming languages and the commonality of bug patterns, we need to include \textit{Unsafely Uncomputation} in the scope of our present study.

\subsection{A Catalog of Bug Patterns in Qiskit}
Quantum programming introduces new quantum-aware bug patterns that differ from existing classical bug patterns. These quantum bug patterns should be identified, and a catalog for these patterns should be presented. 
Due to space limitation, however, we cannot explain more bug patterns in this paper, and in Table~\ref{table:catalog} we list the bug patterns in Qiskit we identified, including those detail described in this section.
To classify the bug patterns listed, we summarize the description for each pattern by pattern name, category, symptoms, causes, and cures \& preventions. Note that this is just a preliminary list of bug patterns in Qiskit, and more bug patterns will be added to the list as we get some new progress.
In the current bug pattern catalog in Table~\ref{table:catalog}, we classify these bug patterns by initialization (1), gate operation (2), measurement (3), and deallocation (4).

\section{Related work}
\label{sec:related-work}

The previous research on bug patterns is mainly focused on classical programming languages. Allen~\cite{allen2002bug} summarizes more than 14 bug pattern categories in Java. Following Allen's work, Hovemeyer and Pugh~\cite{hovemeyer2004finding} present a novel syntactic pattern matching approach to detecting the bug patterns in Java and implement a bug-finding tool called FindBugs~\cite{ayewah2010google}. 
%
Zhang and Zhao~\cite{zhang2007identifying} and Shen~\cite{shen2008xfindbugs} present some bug patterns for AspectJ and develop a tool called {\it XFindBugs} to detect bug patterns in AspectJ.
Our work extends the bug patterns research and identification to the quantum programming languages using the Qiskit language. The bug patterns presented in this paper are different in nature from the existing bug patterns in classical programming languages because they explicitly involve the quantum programming language features such as superposition, entanglement, and no-cloning.

Huang and Martonosi~\cite{huang2018qdb,huang2019statistical} study the bugs for special quantum programs to support quantum software debugging. Based on the experiences of implementing several quantum algorithms, several types of bugs specific to quantum computing are identified. These bugs include incorrect quantum initial values, incorrect operations and transformations, the incorrect composition of operations using iteration, recursion, or mirroring, incorrect classical input parameters, and incorrect deallocation of qubits. The defense strategy for each bug type is also proposed, which mainly uses some assertions to detect the bugs in runtime. While Huang and Matonosi's work targets quantum software debugging, which mainly involves the runtime execution of a program, our work targets identifying the bug patterns to support bug detection through static analysis, which may not need to execute the program and therefore could be more efficient. 

\section{Concluding Remarks}
\label{sec:conclusion}

This paper has identified some bug patterns in the quantum programming language Qiskit to provide both researchers and programmers a clear view of what kind of bugs may happen in quantum programs and how to detect them. The study of bug patterns mainly focuses on bug pattern symptoms, root causes, and cures and preventions. These bug patterns are the first result of our research and do not use every possible quantum-related construct or cover all characteristics of a quantum programming language. New research should cover other remaining quantum-related constructs, as well as the interactions between them.

In our future work, we plan to develop further our approach to investigating more bug patterns in Qiskit programs. We would also like to develop a bug detecting tool based on the identified bug patterns in this paper to support bug finding in Qiskit programs.

\section*{Acknowledgment}
We are grateful to the anonymous reviewers for their suggestions to improve the paper and to Shuhan Lan and Zhongtao Miao for their valuable discussions.

\bibliographystyle{IEEEtran}
\bibliography{IEEEabrv,mybibfile}
\end{document}